# Reversible Modification of Rashba States in Topological Insulators at Room Temperature by Edge Functionalization

Wonhee Ko[1†*], Seoung-Hun Kang[2,3,4†], Qiangsheng Lu[2], An-Hsi Chen[2], Gyula Eres[2], Ho Nyung Lee[2], Young-Kyun Kwon[3,4], Robert G. Moore[2], Mina Yoon[2*], Matthew Brahlek[2*]

[1]Department of Physics and Astronomy, The University of Tennessee, Knoxville, Tennessee 37996, USA

[2]Materials Science and Technology Division, Oak Ridge National Laboratory, Oak Ridge, Tennessee 37831, USA

[3]Department of Information Display, Kyung Hee University, Seoul 02447, Korea

[4]Department of Physics, and Research Institute for Basic Sciences, Kyung Hee University, Seoul 02447, Korea

[†]These authors contributed equally to this work

Email: wko@utk.edu, myoon@ornl.gov, brahlekm@ornl.gov

**Abstract**

Quantum materials with novel spin textures from strong spin-orbit coupling (SOC) are essential components for a wide array of proposed spintronic devices. Topological insulators have necessary strong SOC that imposes a unique spin texture on topological states and Rashba states that arise on the boundary, but there is no established methodology to control the spin texture reversibly. Here, we demonstrate that functionalizing $Bi_2Se_3$ films by altering the step-edge termination directly changes the strength of SOC and thereby modifies the Rashba strength of 1D edge states. Scanning tunneling microscopy/spectroscopy shows that these Rashba edge states arise and subsequently vanish through the Se functionalization and reduction process of the step edges. The observations are corroborated by density functional theory calculations, which show that a subtle chemical change of edge termination fundamentally alters the underlying electronic structure. Importantly, we experimentally demonstrated fully reversible and repeatable switching of Rashba edge states across multiple cycles at room temperature. The results imply Se functionalization as a practical method to control SOC and spin texture of quantum states in topological insulators.

1. Introduction

Topological insulators are bulk insulators characterized by an inverted band structure primarily induced by strong spin-orbit coupling (SOC).[1, 2] This band inversion establishes a non-trivial topological order that guarantees the existence of robust surface states protected by time-reversal symmetry. Moreover, the strong SOC in these materials also leads to the formation of Rashba states at the surface or edges when inversion symmetry is broken.[3, 4] Both topological surface states and Rashba states exhibit spin texture from spin-momentum locking, i.e., each momentum state is associated with a unique spin direction. The unique spin texture of these states facilitates efficient conversion between charge and spin, which makes topological insulators promising for spintronic device applications.[5-12]

Although these advantageous properties make topological materials of general interest for spintronic devices, practical applications require control over the spin texture as well as energy scales for which these phenomena can be observed up to ambient conditions near or above room temperature.[13] From the atomistic perspective, the intrinsic strength of SOC is determined by elemental composition and lattice geometry, making it difficult to modulate without altering the material itself. Traditional approaches for tuning SOC often involve modifying either the gross material composition of the bulk or its underlying lattice structure,[14] which is mostly irreversible. Therefore, finding sensitive tuning "knobs" for the strength of SOC, and thus overall spin texture, that is both reversible and apparent at room temperature is highly desirable for the study of topological insulators as well as future spintronic applications.

Here, we demonstrate that the functionalization of 1D step edges in topological insulators is a highly sensitive platform that enables tuning SOC and thereby switching emergent Rashba edge states reversibly through the edge termination. We investigated 10 QL $Bi_2Se_3$ films grown by molecular beam epitaxy (MBE) (see Methods for detailed growth conditions), which is a prototypical topological insulator.[15, 16] Because the films are grown under the excessive Se flux, step edges are terminated by Se atoms and manifest Rashba edge states in scanning tunneling microscopy/spectroscopy (STM/S) study at room temperature as significantly enhanced d$I$/d$V$ signal along the step edges.[17-19] However, further annealing in the UHV environment (< $10^{-9}$ mbar) drives a fundamental change to the edge state as seen in d$I$/d$V$ spectroscopy at room temperature, where the signal along the step edges disappears. Subsequent reannealing of the films in Se flux fully recovers the Rashba edge states signal. Repeating these two processes of annealing in UHV and annealing in Se flux suppressed and fully recovered the edge states each time, demonstrating complete reversibility. To elucidate the origin of the switching behavior, we employed tight-binding models parameterized by density functional theory (DFT) calculations of the step edges with different terminations. The Se-terminated edges display localized edge states with large Rashba interaction strength, but the Bi-terminated edges show Rashba edge states pushed into the bulk valence band with significantly reduced Rashba interaction strength. The results experimentally and theoretically demonstrate that Se functionalization of step edges switches Rashba edges states on and off, which is thus a promising method to control SOC and spin texture in topological insulators.

## 2. Results and Discussion

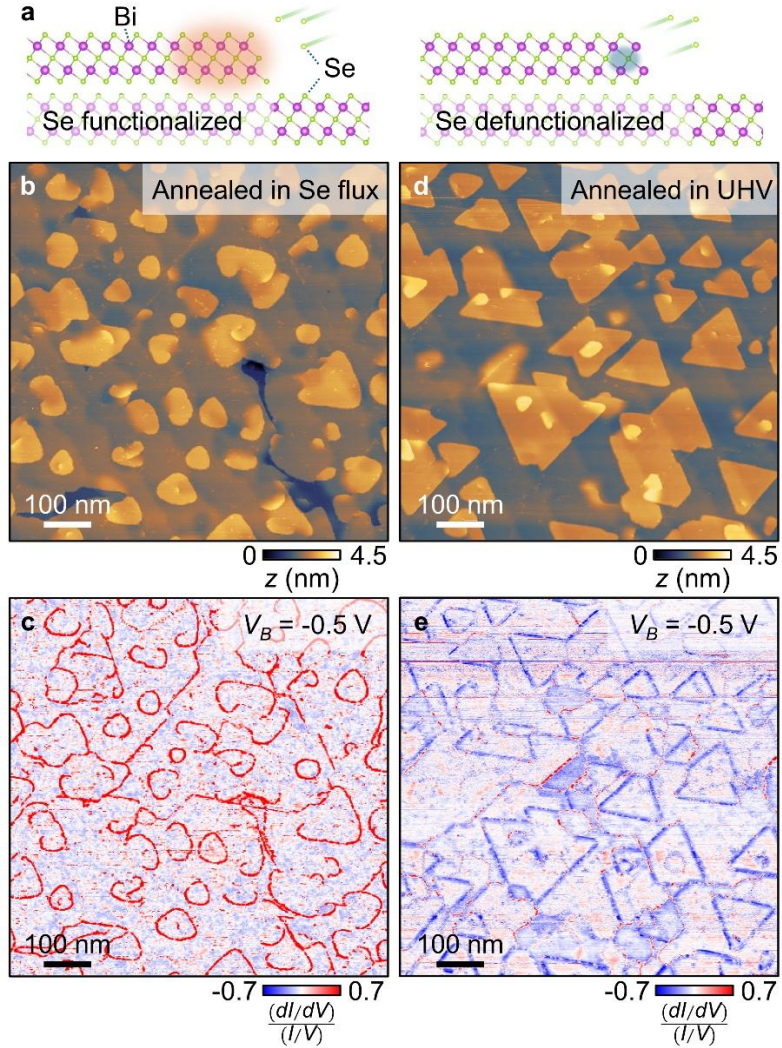

**Figure 1. Rashba edge states in Bi$_2$Se$_3$ depending on functionalization with Se atoms.** a) (left) Schematic of the Bi$_2$Se$_3$ edge functionalization with Se atoms and strongly enhanced Rashba edge states, and (right) schematic of the Bi$_2$Se$_3$ edge defunctionalization by removing Se atoms and suppressed Rashba edge states. b) Large-scale topographic images of the 10 QL Bi$_2$Se$_3$ film after annealing in Se flux ($V_B$ = -0.5 V, $I$ = 10 pA). c) Normalized d$I$/d$V$ maps at $V_B$ = -0.5 V simultaneously taken with the topographic image in (b). d) Large-scale topographic images of the 10 QL Bi$_2$Se$_3$ film after annealing in UHV environment ($V_B$ = -0.5 V, $I$ = 5 pA). e) Normalized d$I$/d$V$ maps at $V_B$ = -0.5 V simultaneously taken with the topographic image in (d).

**Figures 1a** show the schematic of the $Bi_2Se_3$ edge functionalization and defunctionalization, respectively. Here, edge functionalization specifically means forming Se-terminated step edges, which is experimentally realized either by initially growing $Bi_2Se_3$ films with excessive Se (ten times of Bi flux), or by post-annealing the films at 200ºC with a comparable amount of Se flux as used during the growth for about an hour. Edge defunctionalization, in opposition, denotes a process of forming Bi-terminated step edges by detaching Se atoms, which is experimentally achieved by annealing the films at 200ºC with no Se flux in UHV ($< 10^{-9}$ mbar) for about an hour. When terminated with Se atoms, the step edges display significantly enhanced d$I$/d$V$ signal due to the emergence of Rashba edge states, while desorbing those terminating Se atoms suppresses d$I$/d$V$ signal from Rashba edge states. Figure 1b shows a large-scale topographic image of 10 QL $Bi_2Se_3$ film functionalized by Se atoms. The image displays a large flat terrace with islands mostly 1 QL high (~ 1nm), which indicates high-quality layer-by-layer growth of $Bi_2Se_3$ films by MBE. Figure 1c shows the d$I$/d$V$ maps at $V_B$ = -0.5 V taken simultaneously with Fig. 1b, which clearly shows the strongly enhanced d$I$/d$V$ along all step edges. This is consistent with the Rashba edge states developed in $Bi_2Se_3$ due to the strong SOC.[19] Figure 1d shows a large-scale topographic image of the film annealed in UHV condition, which displays similar atomic morphology of 1 QL high islands with identical step height compared to the functionalized film in Fig. 1b. However, the d$I$/d$V$ maps at $V_B$ = -0.5 V show a striking difference (Fig. 1e), where the contrast in d$I$/d$V$ associated with the Rashba edge states has disappeared. The observation demonstrates that the edge states are sensitive to the Se content at the edge termination. Since all the STM measurements were performed at room temperature, our observation shows that the energy scale of Rashba edge states is large enough for switching them on and off in ambient conditions.

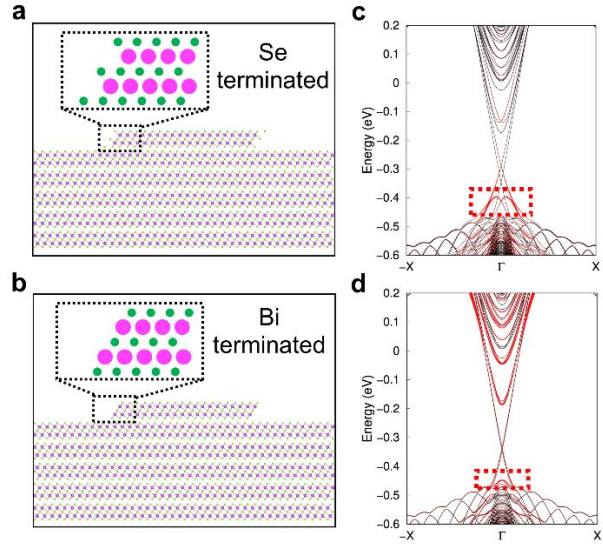

**Figure 2. Theoretical electronic structures of Bi$_2$Se$_3$ films with different edge termination.** a,b) Atomic structure of a 5 QL film with step edges terminated by Se and Bi atoms that represents Se functionalized and defunctionalized edges, respectively. c,d) Band structure of (a) and (b), respectively. The red color is proportional to the band projection at the step edge. The red dotted boxes mark the energy and momentum range where Rashba edge states appear, which show large enhancement of Rashba splitting for the Se-terminated case compared to the Bi-terminated one.

To investigate how Se functionalization affects the edge states, we studied the electronic structures using the tight-binding model based on DFT for a 5 QL Bi$_2$Se$_3$ film with a pair of step edges on top. The step edges can be terminated in several different atomic configurations depending on the Se content (Figure S1, Supporting Information).[20] **Figures 2a,b** show atomic structures with the most Se rich case and Se poor case, and Figs. 2c,d show their band structure, respectively. The bulk band and topological surface states exist for both edge terminations with a similar shape. However, the Rashba edge states below the Dirac point (red boxes in Figs. 2c,d) change drastically with the edge termination. The Se-terminated edge shows an additional band confined to the edge with large Rashba splitting, while the Bi-terminated edge has those bands shifted toward the valence bulk band and much smaller Rashba splitting. In general, the appearance of the Rashba band and Rashba splitting increases with the amount of Se at the edge due to the enhanced electric potential gradient that induces inversion symmetry breaking between the top and bottom QL (Figure S1, Supporting Information). Quantitatively, the Rashba strength $α_R$ and Rashba splitting energy $E_R$ of edge states increase with the Se composition at the edge (Table S1, Supporting Information), which substantiates tunability of Rashba parameters by edge functionalization. We note that the increased electric potential gradient by Se termination also induces 60-meV split between the Dirac points of top and bottom topological surface states in Fig. 2c, which is much smaller for the Bi-terminated case in Fig. 2d. The theoretical results are consistent with the experiment where Se-functionalized edges display a strong edge state signal while defunctionalized edges do not.

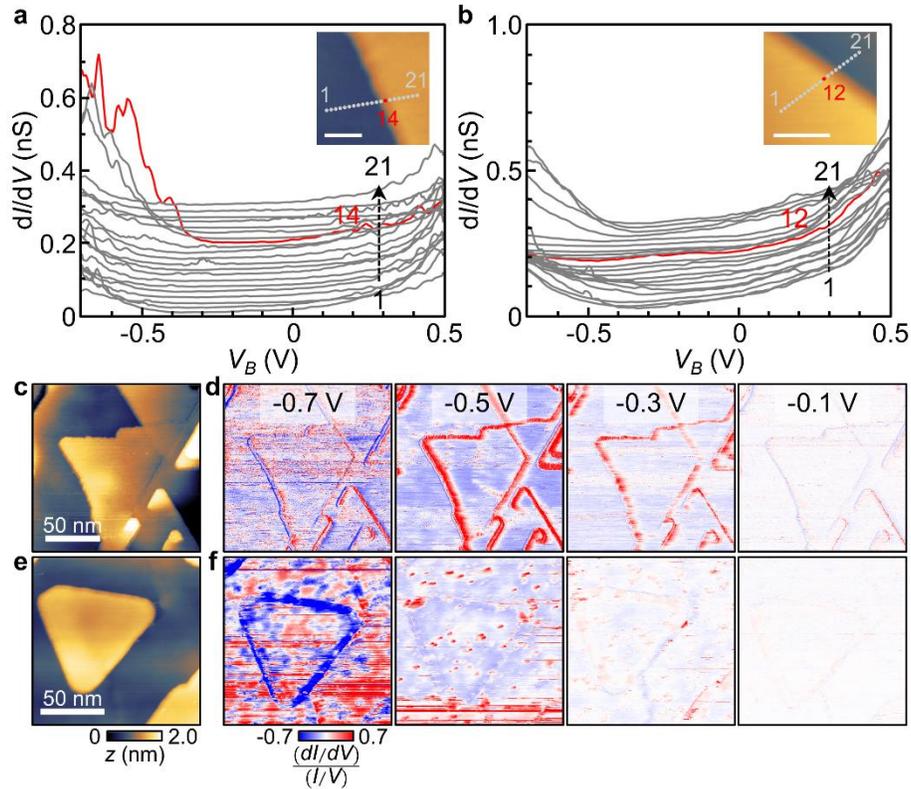

**Figure 3. d$I$/d$V$ spectroscopy at the Se functionalized and defunctionalized step edges.** a,b) d$I$/d$V$ spectra across the step edges functionalized and defunctionalized by Se atoms, respectively. The d$I$/d$V$

curves are offset for clarity. c) Topographic image of $Bi_2Se_3$ film with Se functionalized edges. d) d$I$/d$V$ maps for $V_B$ = -0.7 ~ -0.1 V taken at the same area of (c). e) Topographic image of $Bi_2Se_3$ film with Se defunctionalized edges. f) d$I$/d$V$ maps for $V_B$ = -0.7 ~ -0.1 V taken at the same area of (e).

The d$I$/d$V$ spectroscopy was taken to measure the LDOS and compare it with the band structures in Fig. 2. **Figures 3a,b** show d$I$/d$V$ spectra taken across the step edges that are functionalized and defunctionalized, respectively. For both cases, the spectra inside the bulk (grey lines) show a typical curve with the minimum around $V_B$ = -0.35 V, which indicates that $Bi_2Se_3$ films possess topological surface states with the Dirac point $E_D$ = -0.35 eV regardless of the Se functionalization.[15] However, the spectra right on top of the step edges are drastically different depending on the functionalization. The spectra across the Se functionalized edge show a large increase in d$I$/d$V$ for $eV_B \leq E_D$ (Fig. 3a), which indicates the existence of localized Rashba edge states as predicted in Fig. 2b. However, the defunctionalized edge shows no increase but slight depression in d$I$/d$V$ for $eV_B \leq E_D$ (Fig. 3b). The observation is consistent with the band calculation that shows Rashba edge states pushed into the bulk valence band (Fig. 2d). The d$I$/d$V$ maps confirm such change in spectroscopy generally applies to all step edges (Fig. 3c-f). Figures 3c and d show the topographic image and d$I$/d$V$ maps of the film annealed in Se flux. When $V_B$ = -0.5 V and -0.3 V, the d$I$/d$V$ signal significantly increases along the step edges due to the Rashba edge states. In contrast, the film annealed in UHV shows a drastic change in edge signal. The topographic image still displays step edges with the same height (Fig. 3e), but d$I$/d$V$ maps did not show any increase in the signal along the step edges but slight depression for $V_B$ = -0.5 V (Fig. 3f). The result demonstrates that the topological surface states remain the same, but the edge states are modified by Se functionalization.

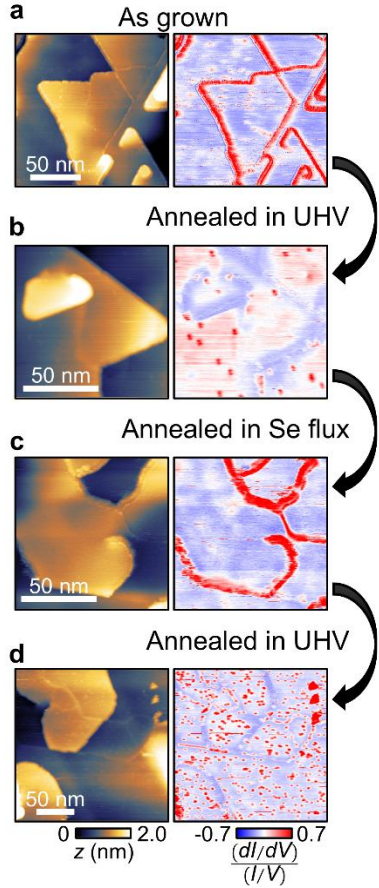

**Figure 4. Reversible modification of Rashba edge states by Se functionalization**. Topographic images and d*I*/d*V* maps for (a) the as-grown film, (b) after annealing in UHV, (c) after annealing in Se flux, and (d) after second annealing in UHV.

Finally, we confirm Se functionalization and defunctionalization are reversible by alternatively annealing in Se flux and UHV (**Fig. 4**). The d*I*/d*V* maps were taken by STM after each step at room temperature. The as-grown film displays Se functionalized edges with Rashba edge states due to the excessive Se flux during the growth (Fig. 4a). Then, annealing the film in UHV defunctionalizes the step edges and removes the Rashba edge states (Fig. 4b). These defunctionalized edges can be functionalized again by annealing in a Se flux environment, which clearly recovers the edge state signal in the d*I*/d*V* map (Fig. 4c). Again, the second annealing in UHV defunctionalizes the edges and removes the Rashba edge states (Fig. 4d). The reversibility of the functionalization at room temperature proves that it is an effective tool to switch on and off the Rashba edge states in ambient conditions.

## 3. Conclusion

In summary, we show that the emergent Rashba edge states that arise at the step edges of $Bi_2Se_3$ can be controlled at room temperature by functionalization and defunctionalization of atomic termination. The functionalization and defunctionalization are achieved by terminating step edges with Se atoms or reducing and changing the termination with Bi, which is experimentally performed by annealing *in situ* in MBE with or without Se flux. This effect is confirmed by using *in situ* STM and DFT calculations, which both highlight that the Rashba edge states are present only for functionalized edges but suppressed for defunctionalized ones. Moreover, DFT calculation suggests that the fundamental driver for this effect is the strength of the SOC dependent on the atomic configuration of the termination, which effectively tunes the energy scale of the Rashba effect at the step edges. Importantly, the process is shown to be fully reversible, i.e., step edges can be functionalized and defunctionalized in turn to repeatedly switch the Rashba edge states on and off. This work highlights the importance of the atomic configuration and detailed geometry of step edges as a subtle tuning knob for local electric potential gradient as well as the effective Rashba interaction strength. Moreover, this effect could be further tailored by, for example, creating 1D wires at steps utilizing surface science knowledge, which could be deployed to integrate other quantum materials that can add additional functionality to these novel edge states, such as superconductors or magnets, or as a route to create a self-aligned 1D gate metal localized at the step boundary.[21, 22] Lastly, our work implies that local electric potential gradient tunes the SOC, and we expect that the same effect can be achieved *ex situ* through an electronic means such as ionic gating and electronic gating with a self-aligned gate structure applied directly to the step edge.[23, 24]

## 4. Experimental Methods

***Sample growth of $Bi_2Se_3$ thin films***

The samples were grown in a home-built molecular beam epitaxy (MBE) reactor at Oak Ridge National Laboratory, which was *in situ* coupled to the STM. The $Bi_2Se_3$ samples were grown on $Al_2O_3$ substrates, which were mounted *ex situ* on the sample holder using silver paste. The substrates were cured at around 150 °C, then cooled and exposed to UV-generated ozone for 10 minutes to clean the surface. The substrates were pumped down and transferred into the MBE reactor, then heated to around 600 °C and exposed to a flux of Se. They were then cooled to about 135 °C where 3 QL of $Bi_2Se_3$ were deposited. The film was then heated to 235 °C where the remainder of the film was grown and then cooled to room temperature and transferred into the STM. The growth was controlled by the amount of Bi, which was calibrated prior to the growth to a flux of $2\times10^{13}$ $cm^{-2}s^{-1}$, and Se was calibrated to about ten times of the Bi flux.

*Scanning tunneling microscopy/spectroscopy*

The STM/STS was taken by Omicron VT-STM operated at ultrahigh vacuum (< $10^{-9}$ torr) and room temperature. The thin film samples grown in the separate MBE chamber were transferred to the STM chamber through the directly connected UHV radial distribution chamber to avoid any exposure to the air. The d$I$/d$V$ spectra were taken by a conventional lock-in technique with a modulation voltage of 50 mV and a modulation frequency of 1 kHz. The d$I$/d$V$ maps were taken in closed-loop mode, where the surface is scanned at a certain bias with the current feedback on while the lock-in modulation voltage is applied to measure d$I$/d$V$ simultaneously.

*Theoretical calculations*

To investigate the electronic properties of the various terminated terrace structures, we used the Slater−Koster type tight-binding (TB) model using hopping parameters in the previous study,[19] which successfully reproduces the DFT band structure for the $Bi_2Se_3$ bulk near the Fermi level. Here, we assumed three *p* orbitals for each of the Bi and Se atoms.

**Acknowledgement**

This work was supported by the UT-Oak Ridge Innovation Institute (UT-ORII) through the UT-ORII SEED grant (W.K., H.N.L.), by the U.S. DOE, Office of Science, National Quantum Information Science Research Centers, Quantum Science Center (S.-H.K., R.G.M.), and by the US Department of Energy, Office of Science, Office of Basic Energy Sciences, Materials Sciences and Engineering Division (G.E., H.N.L., M.Y., M.B.). This work was also partly supported by the Korean government (MSIT) through the National Research Foundation of Korea (NRF) (2022R1A2C1005505) and the Institute for Information & Communications Technology Planning & Evaluation (IITP) (2021-0-01580-001) (Y.-K.K.). This research used resources of the Oak Ridge Leadership Computing Facility at the Oak Ridge National Laboratory, which is supported by the Office of Science of the U.S. Department of Energy under Contract No. DE-AC05-00OR22725.

**Conflict of Interest**

The authors declare no competing interests.